\documentclass[aip,jcp,reprint]{revtex4-1}

\usepackage{amsmath}
\usepackage{amsfonts}
\usepackage{url}
\usepackage{bm}
\usepackage{bbold}
\usepackage{graphics}
\usepackage{paralist}
\newcommand{\bra}{\langle}
\newcommand{\ket}{\rangle}

\begin{document}

\title{An excited-state approach within full configuration interaction quantum Monte Carlo}

\author{N. S. Blunt}
\email{nicksblunt@gmail.com}
\affiliation{University Chemical Laboratory, Lensfield Road, Cambridge, CB2 1EW, U.K.}
\author{Simon D. Smart}
\affiliation{Max Planck Institute for Solid State Research, Heisenbergstra{\ss}e 1, 70569 Stuttgart, Germany}
\author{George H. Booth}
\affiliation{Department of Physics, King's College London, Strand, London WC2R 2LS, U.K.}
\author{Ali Alavi}
\email{a.alavi@fkf.mpg.de}
\affiliation{University Chemical Laboratory, Lensfield Road, Cambridge, CB2 1EW, U.K.}
\affiliation{Max Planck Institute for Solid State Research, Heisenbergstra{\ss}e 1, 70569 Stuttgart, Germany}

\begin{abstract}
We present a new approach to calculate excited states with the full configuration interaction quantum Monte Carlo (FCIQMC) method. The approach uses a Gram-Schmidt procedure, instantaneously applied to the stochastically evolving distributions of walkers, to orthogonalize higher energy states against lower energy ones. It can thus be used to study several of the lowest-energy states of a system within the same symmetry. This additional step is particularly simple and computationally inexpensive, requiring only a small change to the underlying FCIQMC algorithm. No trial wave functions or partitioning of the space is needed. The approach should allow excited states to be studied for systems similar to those accessible to the ground-state method, due to a comparable computational cost. As a first application we consider the carbon dimer in basis sets up to quadruple-zeta quality, and compare to existing results where available.
\end{abstract}

\maketitle

\section{Introduction}
\label{sec:intro}

Quantum Monte Carlo (QMC) methods are a vital tool in the study of many-body systems\cite{Ceperley1979,Kalos1962,Foulkes2001,SorellaSpinLiquid}, regularly providing the most accurate and reliable results for correlated fermionic systems of interest. While their application to the study of ground-state properties is common, the application of QMC methods to the calculation of excited-state properties is more challenging. Such applications typically require the use of an accurate trial wave function with which to constrain the sampling dynamic of the excited state. An accurate trial wave function is generally more difficult to devise, while the results can be particularly sensitive to its form\cite{Schautz2004}, compared to those generated for equivalent ground-state calculations. Yet calculating accurate excited-state properties is key to understanding and predicting various important phenomena in photochemistry and beyond, and therefore much effort continues in this direction.

In 2009, Booth \emph{et al.}\cite{Booth2009} introduced the full configuration interaction quantum Monte Carlo (FCIQMC) method\cite{Booth2009, Spencer2012, Shepherd2012_2, Booth2012}. In common with approaches such as diffusion Monte Carlo (DMC) and Green's function Monte Carlo, FCIQMC is a projector QMC method where the ground-state wave function is sampled by repeatedly projecting out higher energy states with an appropriate stochastically-sampled operator. While DMC performs sampling in real space, FCIQMC samples the wave function in a discrete, antisymmetric space -- typically a basis of Slater determinants. Furthermore, FCIQMC does not require the use of a trial wave function or fixed node approximation to control the fermion sign problem. Rather, the sign problem is controlled by an annihilation step, made efficient by the use of a discrete sampling space.

Already, several approaches have been used to adapt FCIQMC to the calculation of excited states.
Booth and Chan\cite{Booth2012_3} applied a projection operator of the form $\hat{P} = e^{-\Delta\tau^2 (\hat{H}-S)^2}$ to converge to the eigenstate of $\hat{H}$ whose energy is closest to $S$.
Ten-no\cite{Ten-no2013} used the L\"{o}wdin partitioning technique, utilizing an FCIQMC-like dynamic to stochastically evolve the contribution outside the model space and diagonalizing the effective Hamiltonian to obtain multiple eigenvalue estimates.
Humeniuk and Mitri\'{c}\cite{Humeniuk2014} perform an exact diagonalization of the Hamiltonian in a small space containing the majority of the wave function amplitude, and explicitly remove the resulting low-energy states from the projection operator in this diagonalized space.
Recently, three of us have introduced a scheme to project the Hamiltonian into a space of stochastically-sampled Krylov vectors, primarily to consider the calculation of spectral and finite-temperature properties, but also allowing the calculation of individual excited states\cite{Blunt2015_2}. Unlike many traditional excited-state QMC methods, none of these approaches rely on trial wave functions.

In this article we introduce a new approach, whereby FCIQMC wave functions are orthogonalized against each other in order to prevent collapse to the ground state. Multiple FCIQMC simulations are performed simultaneously, one for each state being targeted, and simulations for higher energy states are orthogonalized (by a simple Gram-Schmidt procedure) against those for lower states. This approach is particularly simple, requiring only this single extra step compared to ground-state FCIQMC and very little code to implement in an existing FCIQMC program\cite{Booth2014}. We find that, perhaps surprisingly, the approach is not affected by a noticeable systematic bias in any of our investigations thus far, and therefore systematic improvement to exact results for many excited states is possible.

In section~\ref{sec:fciqmc} we briefly review FCIQMC and its initiator adaptation. In section~\ref{sec:excited_theory} we introduce our approach for obtaining excited states through orthogonalization, including a discussion of the approach's accuracy, and discuss practical details. Results are presented in section~\ref{sec:results}. We study the carbon dimer in cc-pVDZ, cc-pVTZ and cc-pVQZ basis sets, comparing to accurate DMRG results in the quadruple-zeta basis. We also present initiator error convergence for various excited states.

\section{FCIQMC and the initiator adaptation}
\label{sec:fciqmc}

FCIQMC is a projector QMC method whereby the wave function, represented in a sparse form by a list of walkers residing on basis states (usually Slater determinants), is projected to the ground state by repeated application of a projection operator. In FCIQMC the projection operator used is
\begin{equation}
\hat{P} = \mathbb{1} - \Delta\tau (\hat{H} - S \mathbb{1}),
\end{equation}
where $\Delta\tau$ is a small time step, $\hat{H}$ is the Hamiltonian operator and $S$ is a shift parameter, varied slowly to keep the walker population roughly constant.

Therefore, the wave function, $|\Psi(\tau)\ket$ (at imaginary-time $\tau$), obeys
\begin{align}
\Psi(\tau+\Delta\tau) &= \hat{P} |\Psi(\tau)\ket, \\
                      &= [\mathbb{1} - \Delta\tau (\hat{H} - S \mathbb{1})] |\Psi(\tau)\ket.
\label{eq:update_eq}
\end{align}
FCIQMC performs a \emph{stochastic sampling} of this projection so that the correct evolution is achieved on average.

The rationale of this approach can be understood by expanding $|\Psi(\tau)\ket$ in the eigenbasis of the Hamiltonian, $\left\{ |\phi_i\ket ; E_i \right\}$,
\begin{equation}
|\Psi(\tau)\ket = \sum_i c_i(\tau) |\phi_i\ket.
\end{equation}
One finds
\begin{align}
\Psi(\tau+\Delta\tau) &= \sum_i c_i(\tau) [1 - \Delta\tau (E_i - S)] |\phi_i\ket, \\
                      &= \sum_i c_i(\tau) e^{-\Delta\tau (E_i - S)} |\phi_i\ket + \mathcal{O}((\Delta\tau)^2).
\label{eq:energy_decomp}
\end{align}
Thus, excited-state contributions decay relative to the ground state at an exponential rate, and after sufficient applications of $\hat{P}$, a stochastic sampling of the ground-state wave function is achieved.

In practice, the stochastic application of $\hat{P}$ is performed via spawning, death and annihilation steps, which we do not describe here but which have been discussed in detail in previous articles\cite{Booth2009, Spencer2012}.

Because of the sparse sampling of $|\Psi(\tau)\ket$, this approach has significant memory savings compared to traditional FCI. However, the approach cannot use an arbitrarily small amount of memory due to the sign problem. If one considers the population of positive ($n_i^+$) and negative ($n_i^-$) walkers on sites $i$, Spencer $\emph{et al.}$\cite{Spencer2012} showed that a sign problem appears in FCIQMC because, in the absence of annihilation, the desired out-of-phase combination ($n_i^+-n_i^-$) always decays relative to the in-phase combination ($n_i^++n_i^-$), and so at constant populations the desired signal decays to zero (except for rare sign-problem-free systems). This problem is resolved by an annihilation step, whereby two walkers of equal weight and opposite sign on the same determinant cancel out and are removed from the simulation, thus reducing the walker population's growth rate. However, in order to achieve a sufficient annihilation rate a system-specific minimum walker population must be used, in order to reach the so-called `plateau'.

Cleland \emph{et al.} introduced the initiator adaptation to FCIQMC\cite{Cleland2010, Cleland2011} in order to remove this minimum population threshold. In the initiator adaptation, all determinants with more than $n_a$ walkers are dubbed `initiators', with $n_a$ typically equal to $2$ or $3$. The spawning rules are then modified so that non-initiators can only spawn to determinants that were occupied in the previous iteration. Initiators can spawn to any determinant as usual. Previous descriptions of the initiator adaptation also define an exception to the above rules, that non-initiators can spawn to an unoccupied determinant if at least one more spawning occurs to the same determinant with the same sign that iteration, but we note that this exception seems to make no difference to results and so is not a key component.

Because the initiator adaptation greatly reduces the rate of spawning compared to the rate of annihilation, the desired signal ($n_i^+-n_i^-$) no longer decays at small walker populations. As a result, the plateau is removed, and stable simulations (in which the fluctuations in the energy remain bounded in the long imaginary-time limit) can be performed with a relatively small number of walkers.

This adaptation introduces an approximation (and an associated `initiator error') because non-initiators can no longer perform the correct spawning dynamics dictated by Eq.~(\ref{eq:update_eq}), and much of the Hilbert space becomes instantaneously inaccessible. However, the initiator adaptation becomes exact as the walker population increases and more determinants become either initiators or occupied (and therefore can be spawned upon by non-initiators). In the initiator
method, therefore, the number of walkers represents a simulation parameter which can be increased to systematically converge to
exact (FCI) results.  In most cases this limit
can be essentially achieved with substantially fewer walkers than the full FCIQMC scheme requires to overcome the plateau.

Recent work by Petruzielo \emph{et al.}\cite{Petruzielo2012} introduced a semi-stochastic adaptation to FCIQMC, whereby all projection within a subspace (spanned by determinants in a set, $D$) is performed exactly, while the rest of the projection operator is performed stochastically with FCIQMC. This can greatly reduce the stochastic noise associated with the projection, particularly if $D$ is chosen to contain the most highly-weighted determinants. We subsequently performed a detailed investigation of this adaptation in Ref.~(\onlinecite{Blunt2015}), and we use the approach introduced in this latter study in the results of this article. However, this adaptation is not required for the following excited-state approach to work, and the algorithm is unchanged regardless of whether it is in use or not.

\section{Excited-state FCIQMC}
\label{sec:excited_theory}

In our approach to sample excited states, multiple FCIQMC simulations are performed simultaneously, one for each eigenstate to be sampled. We use $m$ and $n$ to label FCIQMC simulations (and therefore energy states), and $i$ and $j$ to label Slater determinants (basis states). To obtain the evolution equation for state $n$, the ground-state evolution equation, Eq.~(\ref{eq:update_eq}), is simply modified so that the components of lower energy states are removed:
\begin{equation}
|\Psi^{n}(\tau+\Delta\tau)\ket = \hat{O}^{n}(\tau+\Delta\tau) [\mathbb{1} - \Delta\tau (\hat{H} - S \mathbb{1})] |\Psi^{n}(\tau)\ket,
\end{equation}
where
\begin{equation}
\hat{O}^{n}(\tau) = \mathbb{1} - \sum_{m < n} \frac{|\Psi^{m}(\tau)\ket \bra \Psi^{m}(\tau)|}{\bra \Psi^{m}(\tau) | \Psi^{m}(\tau) \ket}.
\label{eq:orthog_proj}
\end{equation}
To perform a stochastic version of this evolution for one iteration, each simulation is first evolved using the standard FCIQMC algorithm\cite{Booth2009}. Then, at the end of each iteration, after annihilation has been performed, the overlaps between all pairs of FCIQMC wave functions are calculated, and the components of lower energy FCIQMC states are removed from higher energy ones (and therefore the ground-state simulation is unaffected by this step). Thus, this procedure performs orthogonalization at each iteration against the \emph{instantaneous} FCIQMC wave functions. The rationale behind this step is clear from Eq.~(\ref{eq:energy_decomp}). By removing accurate estimates of low energy states, the simulation converges to the next lowest-energy stationary state of $\hat{H}$ instead.

This step can be performed efficiently by storing the walker weights from all simulations together for a given determinant (i.e., in the same row or column of the array). Some memory wastage is incurred, due to the need to store some zero weights, but this is easily made up for by algorithmic speed and algorithmic simplicity, especially as FCIQMC calculations are not generally memory limited.

In practice, performing the orthogonalization step exactly will introduce some walkers with weights much less than one. For efficiency and memory reasons we stochastically round these small weights to a minimum occupancy threshold, $N_{occ}$, using an unbiased procedure which has been described elsewhere\cite{Overy2014, Blunt2015}. In this article, $N_{occ}=1$. Therefore, FCIQMC simulations may not be exactly orthogonal at the start of each iteration.

We note that a similar approach has been used by Ohtsuka and Nagase\cite{Ohtsuka2010} in their PMC-SD method\cite{Ohtsuka2008}. In this approach higher-energy states are orthogonalized against only a single stochastic wave function estimate, rather than simulating all states simultaneously and orthogonalizing against the differing instantaneous wave functions from each iteration. We expect the approach presented here to be more accurate because we orthogonalize against differing wave functions which will be correct on average, in contrast to orthogonalization against a single approximation to the wave function.

\begin{figure}
\includegraphics{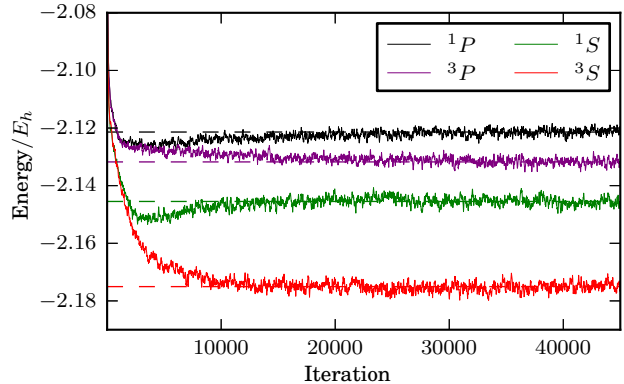}
\caption{An example excited-state FCIQMC calculation, for the He atom in the d-aug-cc-pV5Z basis set. The first four excited states were simulated and $M_s=0$, $L_z=0$ is enforced (thus removing all degeneracy). The ground state was also simulated, but is not shown. $10^4$ walkers were used. The five simulations were started from single determinants in the five available $1s^2$, $1s2s$ and $1s2p$ configurations. Exact values are shown with dashed lines. Initiator and semi-stochastic adaptations were not used.}
\label{fig:He}
\end{figure}

As a basic example for demonstration, Figure~\ref{fig:He} presents results of such a simulation for the He atom in the d-aug-cc-pV5Z basis. Energies for the lowest four excited states are shown converging gradually to the exact values. For this preliminary example, the ground state (${}^1 S$) simulation starts from the $1s^2$ determinant, while ${}^3 S$ and ${}^1 S$ excited simulations start from $1s2s$ determinants and ${}^3 P$ and ${}^1 P$ simulations start from $1s2p$ determinants. As described above, the ground state simulation is evolved using the standard FCIQMC algorithm. All other simulations are evolved concurrently using FCIQMC and, at the end of each iteration, the first excited state (${}^3 S$) simulation is orthogonalized against the ground-state simulation, the second excited state (${}^1 S$) simulation is orthogonalized against the ground and first excited state simulations, and so on, leading to a stable sampling of the appropriate energies.

\subsection{Discussion of accuracy}

Although the rationale behind this procedure is clear, one might wonder how well it will work in practice. A key property of the spawning dynamics in FCIQMC and similar QMC methods is that they are, to a high degree of accuracy\cite{Vigor2015}, unbiased. That is, the expectation value over the probability distribution associated with the possible FCIQMC wave functions at an iteration should obey (up to some normalization constant)
\begin{equation}
E[\boldsymbol{q}] = \boldsymbol{\Psi},
\end{equation}
where $\boldsymbol{q}$ is the stochastic FCIQMC sampling of the exact wave function, $\boldsymbol{\Psi}$. However, functions, $f(\boldsymbol{q})$, which are nonlinear in $\boldsymbol{q}$ should not be estimated by $E[f(\boldsymbol{q})]$ because $E[f(\boldsymbol{q})] \ne f(E[\boldsymbol{q}])$ for nonlinear f. This was found in Ref.~(\onlinecite{Booth2012_2}), where a clear bias was present in reduced density matrices obtained with this approach.

A similar concern can then be raised for the approach taken in the projection step above. On average one wishes to orthogonalize against the \emph{exact} wave function, $\boldsymbol{\Psi}$. However, if $E[\boldsymbol{q}] = \boldsymbol{\Psi}$ then in general
\begin{equation}
E \left[ \: \frac{\boldsymbol{q} \; \boldsymbol{q}^{\dagger}}{\boldsymbol{q}^{\dagger}\boldsymbol{q}} \right] \ne \frac{\boldsymbol{\Psi} \; \boldsymbol{\Psi}^{\dagger}}{\boldsymbol{\Psi}^{\dagger}\boldsymbol{\Psi}},
\label{eq:proj_bias}
\end{equation}
and so it seems that the orthogonalization operator is biased, and using it should introduce a bias towards an incorrect state.

Given the accuracy of the results to follow, one may wonder if there is in fact equality in Eq.~(\ref{eq:proj_bias}), and may seek a proof of this. However, it is simple to devise probability distributions where this is not the case. For example, consider a completely delocalized wave function in a Hilbert space of dimension $D$, where each amplitude in the exact wave function has equal weight, and so $\Psi_i = 1/\sqrt{D}$ for all $i$. Consider an extreme example where this wave function is sampled in an unbiased manner, by placing a single walker of weight one on a site, chosen uniformly. It is straightforward to show that in this case the expectation value in Eq.~(\ref{eq:proj_bias}) is actually proportional to the \emph{identity matrix}. Thus, attempting to orthogonalize with such a poor sampling of $\boldsymbol{\Psi}$ will actually project \emph{any} state towards the zero vector.

However, this is clearly an extreme case. Moreover, in the (admittedly trivial) limit of an entirely single-reference wave function ($\Psi_i = \delta_{ij}$) the bias in this operator tends to zero (by assuming that walkers only reside on site $j$ in this limit). For realistic FCIQMC simulations we will demonstrate that any such bias seems to be extremely small. Indeed, we have so far been unable to find a clear example of such an error in practice. As the bias decreases with walker number, it will appear as a component of the initiator error, and converge along side it. We then only require that the prefactor for any such bias is not excessively large. In all cases to date any bias has been substantially smaller than any measurable initiator error.

To demonstrate the level of accuracy possible for a model system, table~\ref{tab:hubbard} presents results for a 14-site, 1D periodic Hubbard model at $U/t=1$. The sector with total crystal momentum ($\boldsymbol{K}$) and $M_s$ spin quantum number equal to zero was used. The total size of the Hilbert space is $841332$, and approximately $2\times10^4$ walkers were used per state (just sufficient to exceed the plateaus). High accuracy is obtained for all states, and in the absence of initiator error it is clear that any remaining discrepancies are smaller than the magnitude of the smallest random errors which can be realistically achieved within FCIQMC.

It would be interesting to study sign-problem free systems, such as the Heisenberg model on a bipartite lattice, where even smaller walker populations could be used to assess a possible bias, although such systems typically suffer from a severe population control bias\cite{Runge1992, Umrigar1993, Vigor2015} which would need to be removed.

It is hypothesized that this anticipated bias is so small due to the very small contribution of the orthogonalization term each iteration, and the action of the unbiased projection term. Each application of the orthogonalization operator removes all but very small components of the lower-energy states, which already leaves one with accurate estimates. In practice it is found that the overlap between FCIQMC states remains very small throughout the simulations, and therefore the impact of this orthogonalization each iteration is small, and any bias introduced will be small also. This could become a problem if left to grow over many iterations. However, the unbiased part of the projection (as described by Eq.~(\ref{eq:update_eq})) will systematically reduce this error every iteration by projecting back towards the correct states.

\begin{table}
\begin{center}
{\footnotesize
\begin{tabular}{@{\extracolsep{4pt}}cc@{}}
\hline
\hline
FCI & FCIQMC \\
\hline
-14.7147075 & -14.7147079(2) \\
-13.1868974 & -13.1868931(74) \\
-13.1265762 & -13.1265734(63) \\
-12.9714039 & -12.9714105(12) \\
-12.9519154 & -12.952030(264) \\
-12.9252960 & -12.925266(28) \\
\hline
\hline
\end{tabular}
}
\caption{Energy$/t$ for the lowest-energy eigenstates for the 14-site, 1D periodic Hubbard model, at $U/t=1$. Only the $\boldsymbol{K}=\boldsymbol{0}$, $M_s=0$ sector was used. The FCI values are always within 2 standard errors, which are of size $\sim 10^{-4}-10^{-6}t$ for excited states. $1.8\times 10^7$ iterations were performed, with approximately $2\times10^4$ walkers used per state. A CISD space was used to form both the deterministic and trial spaces\cite{Petruzielo2012,Blunt2015}. The initiator adaptation was \emph{not} used, in order to remove initiator error to assess other potential biases instead.}
\label{tab:hubbard}
\end{center}
\end{table}

\subsection{Energy estimators}

In early applications of FCIQMC the energy estimator of choice was the projected Hartree--Fock estimator,
\begin{equation}
E_0 = \frac{\bra D_0 | \hat{H} | \Psi \ket}{\bra D_0 | \Psi \ket},
\label{eq:hf_estimator}
\end{equation}
where $|D_0\ket$ is the Hartree--Fock determinant.

For excited-state energy estimates a trial wave function based estimator is more appropriate, as introduced by Petruzielo \emph{et al.}\cite{Petruzielo2012}, where the Hartree--Fock state is replaced by a multi-determinant trial wave function, $|\Psi_T\ket$. Excited states tend to be more multireference in nature and so single-determinant-based estimates are often poor. In addition to a larger stochastic error, if an inappropriate reference determinant is chosen for the state, then the initiator error can be \emph{substantially} larger. This is due to the wave function being poorly described in sparsely populated regions of the space. The use of a reasonable-quality trial wave function prevents this.

A simple trial wave function for the $n$'th excited state would be the $n$'th excited state obtained from a configuration interaction calculation with singles and doubles (CISD), and this usually works well in practice. Some care must be taken as occasionally states will be ordered differently at the CISD and FCI levels. Therefore a more careful approach is adopted. When performing a simulation for $m$ excited states, one can calculate the first $2m$ excited states at the CISD level. Then the overlap can be taken between each of the $2m$ CISD states and each FCIQMC state, and the best trial wave functions assigned appropriately once convergence has been achieved.

For large production calculations, we obtain better quality trial wave functions using the approach we took in a recent study of semi-stochastic FCIQMC\cite{Blunt2015}. In this, a subspace of dimension $D$ is chosen as the $D$ most populated determinants from the FCIQMC simulation, once convergence is deemed to have been reached. Eigenstates of the Hamiltonian in this subspace are then used to form trial wave functions. This gives better quality trial wave functions than CISD or complete active space (CAS) wave functions in more general systems and geometries.

We note that energy estimates calculated in this way will not be variational. However, variational estimates of the ground state can be obtained by using the estimator $E_{RDM} = \bra \Psi | H | \Psi \ket$, as done in a recent article\cite{Overy2014}. The Hylleraas-Undheim-McDonald theorem\cite{Hylleraas1930, McDonald1933} would allow us to make a similar variational claim for corresponding excited-state estimates, but only if the appropriate states, $\left\{ | \Psi^a \ket \right\}$, obey $\bra \Psi^a | H | \Psi^b \ket = 0$. This will not be true in general because of initiator error. However, we would expect this to be approximately true in most cases. Moreover, although we only consider non-variational estimates in this paper, the trial wave functions used in these expressions will be quite accurate, and so we expect the results to be variational in most cases. This is indeed found to be the case in results presented in this paper (see Figures~\ref{fig:init_VTZ} and \ref{fig:init_VQZ}).

\subsection{Initialization}

The above procedure orthogonalizes against instantaneous FCIQMC wave function estimates, and so no trial wave functions are needed. However, it is not so obvious how to initialize the FCIQMC wave function for each state.

We always choose to start from orthogonal states, such that the first application of the orthogonalization projector in Eq.~\ref{eq:orthog_proj} has no effect. This avoids an initial large drop in walker population, and ensures that the overlap between FCIQMC wave functions then remains small throughout.

\begin{figure}
\includegraphics{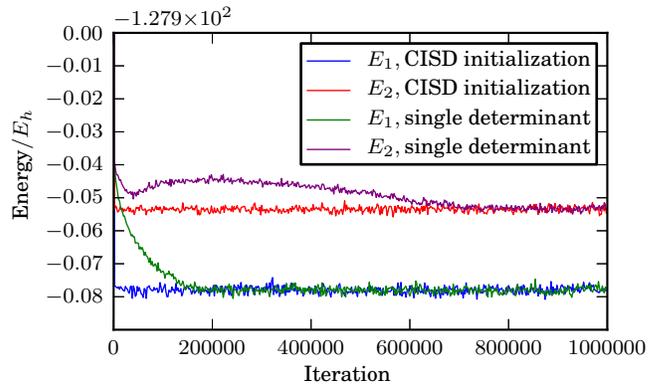}
\caption{Convergence against iteration number for the two lowest excited states (denoted by $E_1$ and $E_2$) for the case of a neon atom in an aug-cc-pVDZ basis, working in the $A_g$ irrep of the $D_{2h}$ point group and with $M_s=0$. All $10$ electrons were correlated. The legend labels denote whether a state was started from a single determinant (by choosing the lowest-energy single excitations of the Hartree--Fock) or from the two lowest excited states from a CISD calculation. Both states are converged in less than $5\times 10^3$ iterations with CISD initialization, compared to almost $10^6$ iterations when using single determinants.}
\label{fig:convergence}
\end{figure}

A simple choice is to start the ground-state wave function from the Hartree--Fock determinant and excited states from the lowest-energy single excitations of this reference. In some situations, such as in Figure~\ref{fig:He}, this is sufficient, but it is not generally true that the dominant determinant in the $n$'th excited is the $n$'th lowest-energy determinant, and in many cases this choice leads to an extremely slow convergence. Figure~\ref{fig:convergence} shows an example of this for the neon atom in an aug-cc-pVDZ basis set, for the two lowest-energy excitations.

Instead, once again a trial wave function is obtained from a subspace calculation (such as CISD or CAS). Even with such basic trial estimates the convergence rate improves substantially. Because these subspaces are usually quite small (the CISD subspace consists of less than $10^6$ determinants in almost all FCIQMC calculations to date), the initial FCIQMC wave functions can be set exactly equal to these trial wave functions, rather than stochastically sampling them. Figure~\ref{fig:convergence} demonstrates the greatly improved convergence for the neon atom by starting from the two lowest excited states from a CISD calculation. Thus, although the approach does not require the use trial wave functions, their use is greatly beneficial. However, it seems that sufficient trial wave functions can often be obtained in a relatively black box manner by using basic subspace techniques.

A final subtlety to mention in the procedure setup arises when the initial wave functions are in the wrong order. It is not uncommon for this to happen at the CAS or truncated CI levels of theory, which we use to generate the initial states. This is a particular problem at stretched geometries due to near degeneracies. Although the orthogonalization step will eventually correct the order of these states, this can take $\sim 10^5-10^6$ iterations. In the results of this article we manually corrected the order of states at initialization, although a more black box approach to automatically perform this reordering should be straightforward.

\section{Results: application to C$_2$}
\label{sec:results}

The carbon dimer is studied as a first application of this approach. This molecule has been studied in detail by both deterministic quantum chemical\cite{Abrams2004, Sherrill2005} as well as QMC methods\cite{Purwanto2009}, including a previous FCIQMC study of the ground state and {\em symmetry-distinct} excited states\cite{Booth2011}, and also by the density matrix renormalization group (DMRG)\cite{Sharma2015}. It is a challenging test for many electronic structure methods due to the multi-reference nature of the ground-state wave function and the presence of low-lying excited states. Indeed, a ground-state crossing exists between $X{}^1\Sigma_g^{{}^+}$ and $B{}^1\Delta_g$ states. This molecule therefore provides a stringent test case. 

In the following, all results correlate $8$ valence electrons. We note that a similar study of C$_2$ has been performed with the auxiliary-field QMC method within their `phaseless' approximation\cite{Purwanto2009}, although the results in that study were performed with all $12$ electrons correlated, and so are not suitable for direct comparison with results presented here.

In all calculations the semi-stochastic adaptation was used with the scheme described in Ref.~(\onlinecite{Blunt2015}), whereby a deterministic space is chosen from the $D$ determinants with the largest weight in the FCIQMC calculation once converged. Here $D$ is the user-specified deterministic space size. In this excited-state context, this approach was modified so that determinants with the largest \emph{summed weight across all simulations} were chosen. The same deterministic space was used for all simulations. It may be more beneficial to use different deterministic spaces for different simulations, but this would require storing multiple deterministic Hamiltonians and would be less efficient, and so we leave this possibility for future investigation.

As described in section~\ref{sec:excited_theory}, the trial space (in which the trial wave functions are generated) was also chosen using this scheme, by choosing the $T$ determinants with the largest summed weight across all simulations. Once again we note that it may be better to choose different trial spaces for different simulations, but that this is computationally less efficient.

Calculations were performed by starting the FCIQMC simulation for state $n$ from the $n$'th lowest-energy state in a subspace calculation. At geometries near equilibrium (which has an internuclear distance of $1.24253\textrm{\AA}$ for C$_2$), a CISD subspace was used for this purpose, whereas a CAS (8,18) subspace was used at highly stretched geometries. In all calculations the time step was varied in the initial iterations to be as large as possible while allowing no (or very few) `bloom' events to occur. A bloom event is defined when more walkers than the initiator threshold are spawned in one event, thus instantly creating an initiator. The presence of many such events typically increases the initiator error and so is undesirable.

\subsection{cc-pVDZ}

\begin{figure}
\includegraphics{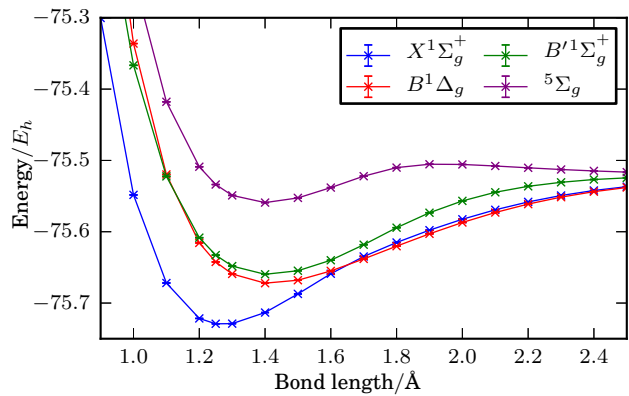}
\caption{Four low-energy states of C$_2$ in a cc-pVDZ basis, obtained from simulations with $M_s=0$ and using the $A_g$ irrep of the $D_{2h}$ point group. Simulations used $10^6$ walkers per state and typically took a deterministic space of size $D=10^4$ and a trial space of size $T=2\times 10^3$. Error bars are plotted but are too small to be visible, typically of order $10^{-7}-10^{-5}\textrm{E}_h$, with initiator error expected to be of the order $10^{-4}\textrm{E}_h$. As expected, a crossing occurs between $X{}^1\Sigma_g^{{}^+}$ and $B{}^1\Delta_g$ states at an internuclear distance of $1.6-1.7$\AA, in addition to a crossing between $B{}^1\Delta_g$ and $B^{\prime}{}^1\Sigma_g^{{}^+}$}
\label{fig:C2_VDZ}
\end{figure}

\begin{figure*}
\includegraphics{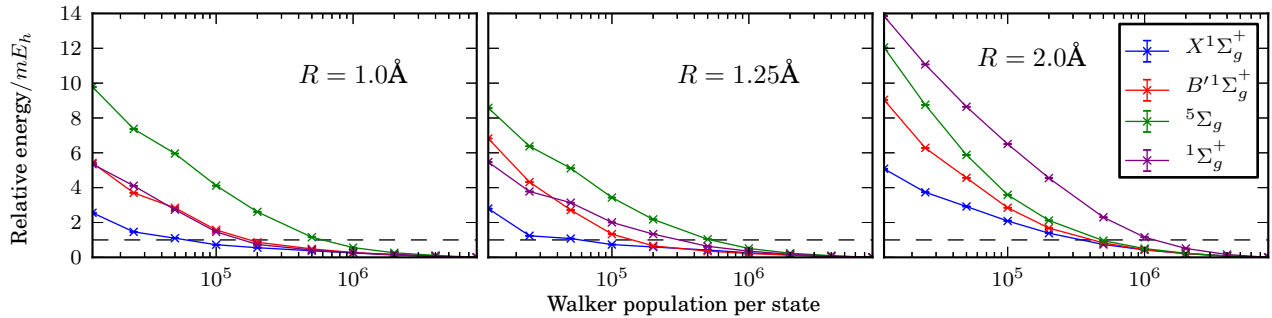}
\caption{Left: $R=1.0$\AA. Center: $R=1.25$\AA. Right: $R=2.0\textrm{\AA}$ for C$_2$ in a cc-pVTZ basis. Initiator error convergence for four low-energy states obtained from simulations with $M_s=0$, $S=\textrm{even}$ and restricted to $\Sigma_g$. Energies are plotted relative to the energy obtained at the largest walker population, $8\times 10^6$ walkers. Simulations used a deterministic space of size $D=10^4$ and a trial space of size $T=2\times 10^3$. Error bars are plotted but are too small to be visible. It is seen in each case that the ground-state initiator error is smaller than excited-state initiator error, although excited-state error is not significantly more difficult to converge. Initiator error is largest in the stretched geometry. The dashed line shows $1\textrm{mE}_h$, from which it is seen that $1\textrm{mE}_h$ accuracy is achieved with roughly $10^6$ walkers in the most challenging case.}
\label{fig:init_VTZ}
\end{figure*}

\begin{figure}
\includegraphics{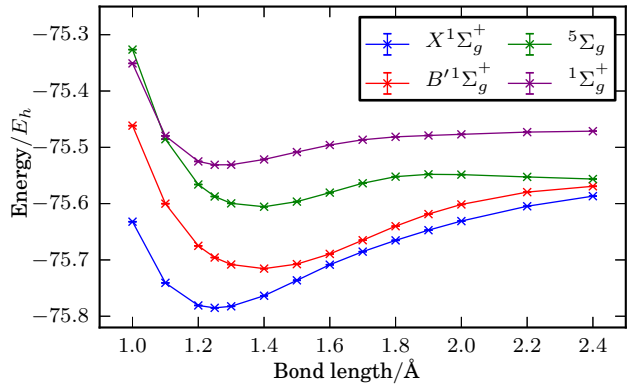}
\caption{Four low-energy states of C$_2$ in a cc-pVTZ basis, obtained from simulations with $M_s=0$, $S=\textrm{even}$ and restricted to $\Sigma_g$. Simulations used $4\times 10^6$ or $8\times 10^6$ walkers per state, with a deterministic space of size $D=10^4$ and a trial space of size $T=2\times 10^3$. Error bars are plotted but are too small to be visible, typically of order $10^{-7}-10^{-5}\textrm{E}_h$, whereas initiator error is expected to be of the order $10^{-4}\textrm{E}_h$. Due to explicit restriction to $L_z=0$ states, the $B{}^1\Delta_g$ state is not present, thus removing a crossing with $X{}^1\Sigma_g^{{}^+}$ seen in figure~\ref{fig:C2_VDZ}.}
\label{fig:C2_VTZ}
\end{figure}

Firstly the use of a double-zeta quality basis set, cc-pVDZ, is considered. Calculations used only those determinants with $M_s=0$ and belonging to the $A_g$ irreducible representation (irrep) of the $D_{2h}$ point group, but no restrictions were placed on the total spin, $S$, or on the angular momentum quantum numbers, $M_l$ and $L$. As such, both $X{}^1\Sigma_g^{{}^+}$ and $B{}^1\Delta_g$ states are present in the calculations.

For each bond length, $10^6$ walkers were used \emph{per state}. The Hilbert space has size $\sim 3 \times 10^{7}$, and so is undersampled, though not substantially. The deterministic space size was usually $D=10^4$. All calculations were performed for $10^6$ iterations, typically enough to perform an accurate error analysis using the `blocking' method\cite{Flyvbjerg1989}. The trial space size was typically taken as $T=2\times 10^3$. These spaces were generated once each simulation had largely converged, taken somewhat arbitrarily at $10^4$ iterations near equilibrium geometry and as $3\times 10^4$ at stretched geometries.

Results are presented in figure~\ref{fig:C2_VDZ}. Based on similar calculations with smaller walker populations, we expect an initiator error of less than $1\textrm{mE}_h$ for all states and for most of the binding curve. Each simulation obtained the $5$ lowest states but only $4$ are presented, with the remaining simulation representing different states at different geometries, due to crossings. The $4$ states present are $X{}^1\Sigma_g^{{}^+}$, $B{}^1\Delta_g$ and $B'{}^1\Sigma_g^{{}^+}$, which have been studied in previous investigations of C$_2$, and also a quintet state. As observed in previous experimental\cite{Martin1992} and theoretical\cite{Abrams2004,Purwanto2009} investigations, a crossing occurs between $X{}^1\Sigma_g^{{}^+}$ and $B{}^1\Delta_g$ states, and also between $B{}^1\Delta_g$ and $B'{}^1\Sigma_g^{{}^+}$ states, all of which are correctly resolved within the approach.

\subsection{cc-pVTZ}

\begin{figure*}[t]
\includegraphics{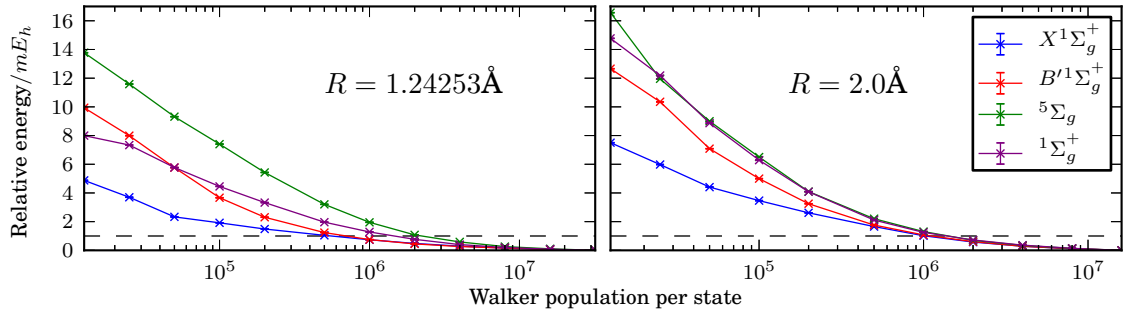}
\caption{Left: $R=1.24253$\AA. Right: $R=2.0\textrm{\AA}$ for C$_2$ in a cc-pVQZ basis. Initiator error convergence for four low-energy states obtained from simulations with $M_s=0$, $S=\textrm{even}$ and restricted to $\Sigma_g$. Energies are plotted relative to the energy obtained at the largest walker population, $3.2\times 10^7$ walkers for $R=1.24253$\AA and $1.6\times 10^7$ walkers for $R=2.0$\AA. Simulations used a deterministic space of size $D=10^4$ and a trial space of size $T=2\times 10^3$. Error bars are plotted but are too small to be visible. As for the equivalent cc-pVTZ plot, the initiator error is larger for excited states than for the ground state. The initiator error is larger in the stretched regime for all states. The dashed line shows $1\textrm{mE}_h$, from which it is seen that $1\textrm{mE}_h$ accuracy is achieved with less than $2\times10^6$ walkers in the most challenging case.}
\label{fig:init_VQZ}
\end{figure*}

\begin{figure}
\includegraphics{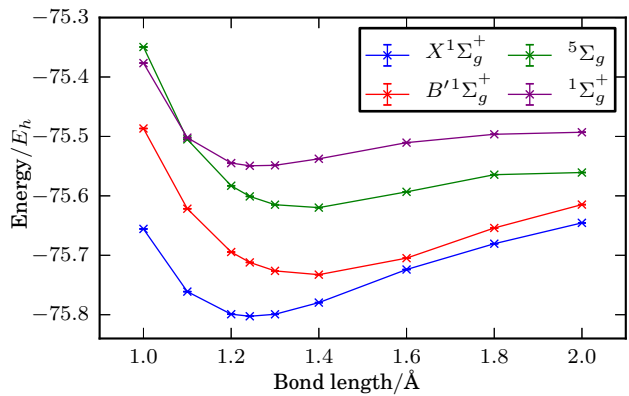}
\caption{Four low-energy states of C$_2$ in a cc-pVQZ basis, obtained from simulations with $M_s=0$, $S=\textrm{even}$ and restricted to $\Sigma_g$. Simulations used $1.6\times 10^7$ walkers per state, with a deterministic space of size $D=10^4$ and a trial space of size $T=2\times 10^3$. Error bars are plotted but are too small to be visible, typically of order $10^{-6}-10^{-5}\textrm{E}_h$, whereas initiator error is expected to be of order $10^{-4}\textrm{E}_h$.}
\label{fig:C2_VQZ}
\end{figure}

Next, the use of a cc-pVTZ basis is considered. Due to difficulties caused by near degeneracies, a more restricted symmetry subspace is used. Firstly, molecular orbitals which are eigenfunctions of the $\hat{L}_z$ operator (z-component of angular momentum) are used, and simulations are restricted to $L_z=0$. This removes the crossing between $X{}^1\Sigma_g^{{}^+}$ and $B{}^1\Delta_g$ states. Secondly, instead of using determinants as basis states, time-reversal symmetrized functions\cite{Smeyers1973} are used instead. These allow the total spin, $S$, to be even or odd, and so the even-$S$ sector is considered to remove all triplet states (but \emph{not} quintet states). Finally, we restrict to states which are even with respect to inversion symmetry ($\Sigma_g$). With these symmetry restrictions, the total space size is $\sim 5\times 10^9$. In each simulation, $D=10^4$ and $T=2\times10^3$. Once again, $5$ states were obtained in each simulation and $4$ are presented here, with the remaining state differing due to crossings. With the $B{}^1\Delta_g$ state no longer in the space considered, the additional state considered is of ${}^1\Sigma_g^{{}^+}$ nature.

Figure~\ref{fig:init_VTZ} presents the initiator error convergence of the four states considered. This figure shows the energies at each population, relative to the corresponding energies at $8 \times 10^6$ walkers (per state), thus giving an accurate estimate of the initiator error throughout. The left subplot shows a compressed geometry ($R=1.0$\AA), the center plots shows equilibrium geometry ($R=1.25$\AA) and the right plot shows a stretched geometry ($R=2.0$\AA). For each geometry it is clear that the ground state has the smallest initiator error. This is perhaps expected, because excited states tend to be more multi-reference in nature. 
However, the difference in initiator error between ground and excited states is not too great, and all initiator curves seem to converge to much better than milli-Hartree accuracy with $8\times 10^6$ walkers. The dashed line shows $1\textrm{mE}_h$ accuracy, from which it is seen that $1\textrm{mE}_h$ accuracy is achieved with roughly $10^6$ walkers in the most difficult case (with the FCI space size being $\sim 5\times 10^9$). It is also seen that stretched geometries are the most challenging. This is consistent with the results of Ref.~(\onlinecite{Abrams2004}), where larger nonparallelity errors were found for various quantum chemical methods, with the largest errors for CISD occurring for large bond lengths (in the case of an RHF reference, as is considered here).

Other sources of error, besides initiator error, could be present in figure~\ref{fig:init_VTZ}. In particular, the bias described in section~\ref{sec:excited_theory} due to the nonlinear projection. However, given that the ground-state energy (which contains no such bias, as no orthogonalization is performed for this simulation) is of a similar size, and based on observations of the accuracy of the orthogonalization approach in calculations without the initiator approximation, we believe that any such error is very small.

Binding energy curves are presented in figure~\ref{fig:C2_VTZ}. Based on the results of figure~\ref{fig:init_VTZ}, a population of $4\times 10^6$ walkers was used per state for all geometries except the three most stretched cases ($R=2.0$\AA, $R=2.2$\AA, $R=2.4$\AA), where $8\times 10^6$ walkers were used. Figure~\ref{fig:init_VTZ} suggests this should give substantially sub-milli-Hartree accuracy in all cases.  All other simulations parameters were identical to those used to produce figure~\ref{fig:C2_VTZ}, as defined above.

\subsection{cc-pVQZ}
\label{sec:VQZ}

A cc-pVQZ basis set is also investigated, in order to compare to a recent high-accuracy DMRG study of C$_2$\cite{Sharma2015} in the same basis. The space size for this system is $\sim 6\times 10^{11}$ (using the same spin and symmetry restrictions as for the cc-pVTZ calculations). We again choose $D=10^4$ and $T=2\times10^3$ for all results. Figure~\ref{fig:init_VQZ} presents initiator error convergence at equilibrium ($R=1.24253$\AA) and stretched ($R=2.0$\AA) geometries, using up to $3.2 \times 10^7$ and $1.6 \times 10^7$ walkers per state in the two cases, respectively. Once again, the dashed line shows $\textrm{1mE}_h$ accuracy, and it is seen that this can be achieved with less than $2\times10^6$ walkers in all cases. By $8\times 10^6$ walkers, errors appear to be much smaller than $\textrm{1mE}_h$. Once again, and as expected, initiator error is smallest for the ground state, and initiator error is larger in the stretched regime.

\begin{table*}[t]
\begin{center}
{\footnotesize
\begin{tabular}{@{\extracolsep{4pt}}lcccccc@{}}
\hline
\hline
$R$/\AA & \multicolumn{3}{c}{DMRG energies} & \multicolumn{3}{c}{FCIQMC energies} \\
\cline{1-1} \cline{2-4} \cline{5-7}
$1.0$    & - & - & - & $-0.655705(1)$ & $-0.486651(6)$  & $-0.376542(12)$ \\
$1.1$    & $-0.76124$ & $-0.62183$ & $-0.50228$ & $-0.761142(1)$ & $-0.621703(4)$  & $-0.502117(4)$  \\
$1.2$    & $-0.79920$ & $-0.69459$ & $-0.54490$ & $-0.799132(2)$ & $-0.694501(5)$  & $-0.544788(4)$  \\
$1.24253$ & $-0.80264$ & $-0.71208$ & $-0.54953$ & $-0.802575(3)$ & $-0.711995(13)$ & $-0.549421(6)$  \\
$1.3$    & $-0.79933$ & $-0.72633$ & $-0.54871$ & $-0.799271(2)$ & $-0.726263(3)$  & $-0.548611(2)$  \\
$1.4$    & $-0.77965$ & $-0.73267$ & $-0.53776$ & $-0.779606(2)$ & $-0.732612(4)$  & $-0.537660(2)$  \\
$1.6$    & $-0.72401$ & $-0.70487$ & $-0.51054$ & $-0.723949(3)$ & $-0.704805(4)$  & $-0.510472(4)$  \\
$1.8$    & - & - & - & $-0.680562(3)$ & $-0.654071(2)$  & $-0.496394(4)$  \\
$2.0$    & $-0.64552$ & $-0.61469$ & $-0.49290$ & $-0.645482(3)$ & $-0.614701(6)$  & $-0.492973(10)$ \\
\hline
\hline
\end{tabular}
}
\caption{Comparison of DMRG and FCIQMC results for the lowest three states of C$_2$ cc-pVQZ in the ${}^1\Sigma_g^{{}^+}$ irrep. All energies are in Hartrees and are shifted by $+75.0\textrm{E}_h$. Results agree to $\sim 10^{-4}\textrm{E}_h$ accuracy, with the (variational) DMRG results typically being lower by $0.1\textrm{mE}_h$ (note that the quoted FCIQMC errors are \emph{stochastic} error bar sizes, not systematic error). FCIQMC results all use $1.6\times 10^7$ walkers per state and are plotted in figure~\ref{fig:C2_VQZ}. DMRG results are taken from Ref.~(\onlinecite{Sharma2015}).}
\label{tab:DMRG_comparison}
\end{center}
\end{table*}

It is interesting to compare the difference in initiator error between cc-pVTZ and cc-pVQZ. The cc-pVQZ space size is roughly 120 times larger than that for the cc-pVTZ case, yet a much smaller increase in walker number is required for similar accuracy with FCIQMC. For example, in the ground state at equilibrium geometry, $\sim 5\times 10^5$ walkers are needed for $\textrm{1mE}_h$ accuracy in the cc-pVTZ case, compared to $\sim 10^6$ walkers for the cc-pVQZ case (to more accurately describe the \emph{memory} increase, we note that the number of occupied basis states in these two cases were $1.5\times 10^6$ and $3.5\times 10^6$, respectively). This is a demonstration of sub-linear scaling of required memory in the FCIQMC approach, which has been observed previously\cite{Cleland2011,Shepherd2014}.

Table~\ref{tab:DMRG_comparison} presents a comparison between FCIQMC and recently-published DMRG results\cite{Sharma2015}, for the three lowest states in the ${}^1\Sigma_g^{{}^+}$ irrep. By comparing to the accuracy of ground-state DMRG calculations with a larger number of renormalized states, the error in the DMRG excited state results was estimated at less than $0.1\textrm{mE}_h$ for the entire binding curve, with 4000 spin-adapted renormalized states kept for each excited state. Based on the results of figure~\ref{fig:init_VQZ}, $1.6\times 10^7$ walkers were used per state for our FCIQMC calculations. FCIQMC results agree with the DMRG results to a very high degree of accuracy. For most bond lengths, the DMRG results are approximately $0.1\textrm{mE}_h$ below the FCIQMC values, with the DMRG results being variational. This remaining small error in the FCIQMC results is most likely initiator error, and could be removed by using larger walker populations. We again note that it is possible that some error is due to the bias described in section~\ref{sec:excited_theory}, however, we believe this unlikely because of the similar-sized discrepancy compared to DMRG is also present in the ground-state energy, which can contain no such bias.

At $R=2.0$\AA, and for the two excited states presented in table~\ref{tab:DMRG_comparison}, the FCIQMC energies are slightly ($<0.1\textrm{mE}_h$) lower than the DMRG values. This is despite the fact that figure~\ref{fig:init_VQZ} suggests that even the FCIQMC error is larger at this stretched geometry than near equilibrium. This seems to suggest that the DMRG also has larger errors at stretched geometries, although in any case the errors seem to be very small. Results of table~\ref{tab:DMRG_comparison} are plotted in figure~\ref{fig:C2_VQZ} (which also includes a quintet state, not studied in the DMRG calculations of Ref.~(\onlinecite{Sharma2015})).

\section{Discussion}

In chemical systems it is quite common for there to be a substantial gap between the ground and first excited states, and so excited-state contributions typically die away quickly in ground-state FCIQMC simulations. In contrast, excited states typically populate the energy spectrum much more densely, and near-degeneracies are common. Such near-degeneracies lead to very slow convergence and very long autocorrelation lengths, both of which lead to having to perform a large number of iterations. Based on the calculations presented in this paper, we estimate that it is not uncommon to have to perform up to $10$ times more iterations to obtain enough independent samples of the energy from which to draw meaningful averages and error bars for all states, compared to a ground-state calculation. Moreover, since excited-state calculations require one FCIQMC simulation per state, the time required per iteration to obtain $N$ states is approximately $N$ times greater than for a ground-state FCIQMC calculation. Therefore, excited-state FCIQMC calculations are undoubtedly more expensive than ground-state calculations. However, we believe that the difference is not in general prohibitively large, and is helped by efficient parallelization over processing cores\cite{Booth2014} (additional parallelization of the algorithm over excited states is likely to be more efficient, but is not explored here). We therefore expect that most systems for which ground-state FCIQMC is possible will also be suitable for this excited-state extension.

The method is also able to obtain degenerate states with great accuracy. However, in such cases the issue of long auto-correlation lengths is often exacerbated due to the potential for rotation of the state within the degenerate subspace.
This can be ameliorated by removing the degeneracy, if possible, by enforcing a further symmetry, such as $\hat{L}_z$ symmetry. In the same manner as for other FCIQMC calculations, it seems advisable to enforce as many symmetries as possible. This also reduces the possibility of state crossings, as seen in the C$_2$ results above.

\section{Conclusion}

We have introduced a simple approach for obtaining excited states from the FCIQMC method, by performing an orthogonalization step between multiple simultaneous FCIQMC simulations. This orthogonalization is performed against instantaneous stochastic snapshots of the true wave functions, yet this does not prevent the method from obtaining highly accurate results.

Low-lying states of the C$_2$ molecule have been studied in cc-pVXZ basis sets, for $X=2$, $3$ and $4$. In the quadruple-zeta basis set results were compared to accurate DMRG benchmarks, which were reproduced with roughly $\sim 0.1\textrm{mE}_h$ error. It was shown that the initiator adaptation is also effective for excited-state calculations. Although initiator error does tend to be somewhat larger for excited states than for ground states, the favorable sub-linear scaling of walkers with Hilbert space size seems to also hold for the excited states, as well as the ground state, for the C$_2$ case considered here.

It should be straightforward to combine this approach with the recent unbiased reduced density matrix calculations within FCIQMC\cite{Overy2014}, which will allow investigation of other properties of excited molecules, such as dipole moments, dipole polarizabilities, and nuclear forces\cite{Thomas2015_2}.

\section{Acknowledgments}

We thank Alex Thom for useful discussions and Sandeep Sharma for technical assistance.
N.S.B. gratefully acknowledges Trinity College, Cambridge for funding.
G.H.B. gratefully acknowledges the Royal Society for funding via a university research fellowship.
This work has been supported by the EPSRC under grant no. EP/J003867/1.

%

\end{document}